\documentclass[preprint,12pt]{elsarticle}

\usepackage{amssymb}
\usepackage{graphicx}% Include figure files
\usepackage{dcolumn}% Align table columns on decimal point
\usepackage{bm}% bold math
\usepackage{nicefrac}
\usepackage{comment}
\usepackage{amsmath}

\journal{Journal of Computational Physics}

\begin{document}

\begin{frontmatter}



\title{Symplectic Integration of Magnetic Systems}


\author{Stephen D. Webb}
\ead{swebb@txcorp.com}

\address{Tech-X Corporation, 5621 Arapahoe Ave., Boulder, CO 80303}

\begin{abstract}
Dependable numerical results from long-time simulations require stable numerical integration schemes. For Hamiltonian systems, this is achieved with symplectic integrators, which conserve the symplectic condition and exactly solve for the dynamics for an approximate Hamiltonian. The conventional generating function formalism used for Hamiltonian systems is problematic for magnetic systems, as the Hamiltonian is not easily separable. This paper presents a derivation of symplectic integrators from a discretized action. Using this method, it is shown that the integrator due to Boris is a symplectic integrator in the non-relativistic limit, and a relativistic integrator is derived.
\end{abstract}


\end{frontmatter}

% \linenumbers

\section{Introduction}

Symplectic integration \cite{ruth:83, ruth_forest:89, yoshida:90, sanz-serna:88, forest:06} has become a staple of accelerator physics and astrophysics simulations, as they provide unconditional stability, if not the short term accuracy of Runge-Kutta type schemes. While numerical solutions for systems such as magnetized plasmas are not derived directly from Hamiltonian systems, the canonical method due to Boris \cite{boris:70, birdsall_langdon:85, hockney_eastwood:89} shows many of the properties of a symplectic integrator.

The primary difficulty in developing symplectic integrators for magnetic systems is, as was pointed out by Ruth \cite{ruth:83}, is the $\vec{p} \cdot \vec{A}$ term that leads to implicit forms and which is not separable into an exactly integrable Hamiltonian. Even worse is for relativistic systems, where no clear expansion parameter for the radical in the Hamiltonian exists. In this case, the kinetic energy is not even close to separable. This problem does not appear in Lagrangian mechanics, where the vector potential appears in a $\dot{q} \cdot \vec{A}$ form outside the radical. However, Lagrangian mechanics lacks the canonical transformation formalism used in deriving symplectic integrators.

To obtain geometric integrators from a Lagrangian formalism, it is best to approach the problem using a discretized action integral. This method, described in \cite{marsden_west:01} and the citations therein, obtains recursion relations for the $q_k$ in configuration space that conserve the symplectic two-form.

In this paper, we present the formalism necessary to show that the Boris method is a symplectic integrator. An overview of discrete Lagrangian mechanics based on the work of Marsden and West \cite{marsden_west:01} is first presented. This method is then applied to Lagrangians with vector potentials, first the nonrelativistic limit, where this is shown to be the Boris update. Then, the equivalent result is obtained for the relativistic Lagrangian. We conclude with some remarks on higher order integration and the method due to Yoshida~\cite{yoshida:90}, and implications to higher order electromagnetic particle-in-cell algorithms.

\section{Discretized Lagrangian Mechanics}

As discussed by Marsden and West (\cite{marsden_west:01}, and citations therein), the Lagrangian action integral may be approximated by some discretization scheme by
\begin{equation}
\int_0^t L(q, \dot{q}, t') dt' \approx \sum_{k = 0}^{N-1} L_D(q_{k + 1}, q_k, t_k)
\end{equation}
where $t_k = t_0 + k h$ for a time step $h$. Minimizing this action against variations $\delta q_k$ of the physical trajectory, with $\delta q_0 = \delta q_N = 0$ to fix the endpoints, gives a variation of the discrete action
\begin{equation}
\begin{split}
\delta S_D = \sum_{k = 1}^{N - 1} \biggl{(} \frac{\partial}{\partial q_{k + 1}} L_D(q_{k +1}, q_k, t_k) \delta q_{k + 1} + \frac{\partial}{\partial q_k} L_D(q_{k +1}, q_k, t_k) \delta q_k \biggr{)} + \\ \partial_{q_0} L_D(q_0, q_1) \delta q_0 + \partial_{q_N} L_D(q_N, q_{N-1}) \delta q_N = 0
\end{split}
\end{equation}
By shifting the summations to match the indices of the variations, this gives the discrete Euler-Lagrange (DEL) equations
\begin{equation}
D_2 L_D(q_{k + 1}, q_k) + D_1 L_D(q_k, q_{k - 1}) = 0
\end{equation}
where $D_n$ is the derivative with respect to the $n^{th}$ variable. 

In the continuous Lagrangian limit, the symplectic two-form
\begin{equation}
\Omega_L(q, \dot{q}) = \frac{\partial^2 L}{\partial q_i \partial \dot{q}_j} \mathbf{d} q^i \wedge \mathbf{d} q^j + \frac{\partial^2 L}{\partial \dot{q}_i \partial \dot{q}_j} \mathbf{d} \dot{q}^i \wedge \mathbf{d} q^j
\end{equation}
is conserved under the Euler-Lagrange equations. For the discretized Lagrangian, the tangent bundle $T \mathbb{Q}$ does not include $\dot{q}$, as the time derivatives do not appear in the Lagrangian. The discretized symplectic two-form that is conserved is given by
\begin{equation}
\Omega_{L_D}(q_0, q_1) = \frac{\partial^2 L}{\partial q_0^i \partial q_1^j} \mathbf{d} q_0^i \wedge \mathbf{d} q_1^j
\end{equation}
Because $\Omega_{L_D} = \mathbf{d} \Theta_L$ is an exterior derivative, and $\mathbf{d}^2 = 0$, this symplectic two-form is conserved for all solutions of the DEL equations. This proof is provided in detail in Marsden and West.

Because $\dot{q}$ is replaced by a finite difference in the discrete Lagrangian, it is important to note that any velocity-like variables are auxiliary and do not play a role in the underlying geometric structure. This is an important distinction with Hamiltonian symplectic integrators, where the $p$ and $q$ both play a role in the geometry -- the Hamiltonian symplectic two form explicitly involves the momentum.

As an example of how this yields a symplectic integrator, consider the Lagrangian for a one dimensional particle in a potential
\begin{equation}
L(q, \dot{q}, t) = \frac{1}{2} \dot{q}^2 - V(q, t)
\end{equation}
The discretization is non-unique -- indeed choosing $q \mapsto \nicefrac{(q_k + q_{k+1})}{2}$ yields an implicit integration scheme for general $V$, while choosing $q \mapsto q_{k+1}$ will yield the explicit integrator below. Because we are interested in explicit integrators, we will consider the discrete Lagrangian
\begin{equation}
L_D(q_{k+1}, q_k) = \frac{1}{2} \frac{(q_{k+1} - q_k)^2}{h} - V(q_{k+1}, (k+1) h) h
\end{equation}
for a discrete time step $t$. Applying the DEL equations to this discrete Lagrangian yields
\begin{equation}
q_{k +1} - 2 q_k + q_{k - 1} = - \frac{\partial V}{\partial q_{k}} (q_k, kh) h
\end{equation}
which we recognize immediately as $F = m \ddot{q}$ in the form of a central differencing. If we define the velocity vector to be
\begin{equation}
v_{k + 1} = \frac{q_{k + 1} - q_k}{h}
\end{equation}
then this yields the first order symplectic integrator
\begin{equation}
v_{k + 1} = v_k - \frac{\partial V}{\partial q_{k}} (q_k, k t) h
\end{equation}
\begin{equation}
q_{k + 1} = q_k + v_{k + 1} h
\end{equation}
which has the usual first order leapfrog scheme.

Unlike in Hamiltonian symplectic integration schemes, where the generating function defines the $p$ and $q$ update sequences explicitly, we were forced here to introduce the velocity as an auxiliary variable, turning our $N$ second order recursion relations into $2 N$ first order recursion relations. This is because, in Lagrangian mechanics, the $q$ and $\dot{q}$ variables are not independent -- the same reason there are no useful canonical transformations to generate symplectic integrators. This will become a useful construct when we consider the relativistic Lagrangian in Section IV.

To obtain a higher order integrator, it is necessary to sum two or more Lagrangians per time step, in the form
\begin{equation}
S_D = \sum_{k = 0}^{N} \sum_{i = 1}^M L_D^{i, k} (q_k^{i+1}, q_k^i, \gamma^i h)
\end{equation}
where the individual $L_D^i$ are for a substep of a single time step. From here on, we will drop the $k$ indexing of the discretized Lagrangian as we are only concerned with a single time step linking $k$ to $k+1$. By carrying out the same variational calculation, the resulting DEL equations for this decomposed system is given by
\begin{subequations}
\begin{equation}
D_2 L_D^i (q_k^{i - 1}, q_k^i, \gamma^i h) + D_1 L_D^{i+1}(q_k^i, q_k^{i+1}, \gamma^{i + 1} h) = 0
\end{equation}
\begin{equation}
D_2 L_D^M (q_k^{M-1}, q_k^M, \gamma^M h) + D_1 L_D^1(q_{k+1}^0, q_{k+1}^1, \gamma^1 h) = 0
\end{equation}
\end{subequations}
where the $\gamma^i$ index the time sub-steps and the steps are joined by $q_k^M = q_{k+1}^0$. The second equation give the linkage between the last step of one time step, and the first step of the next one -- in practice this frequently manifests as the final velocity of one time step equaling the initial velocity of the next.

If $L_D(q_0, q_1, h) = \sum_{i} L_D^i$ is the discrete Lagrangian, then the adjoint of the Lagrangian is defined as 
\begin{equation}
L_D^*(q_0, q_1, h) \equiv - L_D(q_0, q_1, -h)
\end{equation}
It is straightforward to show that a self-adjoint discrete Lagrangian $L_D^* = L_D$ only contains odd order corrections to the action integral, and therefore any self-adjoint discrete Lagrangian is automatically second order. We will therefore consider only self-adjoint Lagrangians in this paper.

\section{Symplectic Integrators for Magnetic Systems} \label{borisSection}

We will now use this formalism to compute an explicit symplectic integration scheme for Lagrangians with vector potentials, beginning in the non-relativistic regime with a Lagrangian
\begin{equation}
L(q, \dot{q}, t) = \frac{1}{2} m \dot{q}^2 + \dot{q} \cdot \frac{e}{c} \vec{A} (q, t) - e \phi(q, t)
\end{equation}
The dynamics are unchanged by the addition of a total derivative, or by dividing the Lagrangian by a constant. Thus, dividing through by the mass and defining

\begin{equation}
\vec{a} = \frac{e}{m c} \vec{A}
\end{equation}
\begin{equation}
\varphi = \frac{e}{m} \phi
\end{equation}

yields the normalized Lagrangian
\begin{equation}
L(q, \dot{q}, t) = \frac{1}{2} \dot{q}^2 + \dot{q} \cdot \vec{a}(q, t) - \varphi(q, t)
\end{equation}
Given the close relationship between symplectic integration schemes and Trotter splitting for the Schr\"{o}dinger equation, it is perhaps fruitful to take inspiration from path integration. It is well-known (for a discussion of this, see \cite{schulman_text}) that, in evaluating a path integral with gauge fields, it is necessary to evaluate the gauge field either as
\begin{equation} \label{explicit}
\vec{A}(q) \mapsto \frac{1}{2} \left ( \vec{A}(q_k, 0) + \vec{A}(q_{k+1}, h) \right )
\end{equation}
or
\begin{equation} \label{implicit}
\vec{A}(q) \mapsto \vec{A} \left ( \frac{q_{k+1} + q_k}{2}, \nicefrac{h}{2} \right )
\end{equation}
to preserve the gauge invariance of the equations of motion. We will consider only the former, as this will lead us to an explicit algorithm that is second order accurate.

Consider the discretized splitting into two Lagrangians given by
 \begin{subequations}\label{borisSplit}
\begin{equation}
\begin{split}
L_D^1(q_k^0, q_k^1, t) = \frac{1}{2} \frac{(q_k^1 - q_k^0)^2}{\nicefrac{h}{2}} +\dots \\ (q_k^1 - q_k^0) \cdot \frac{1}{2} \left ( \vec{a}(q_k^0, 0) + \vec{a}(q_k^1, \nicefrac{h}{2}) \right ) - \frac{h}{2} \varphi(q_k^1, \nicefrac{h}{2})
\end{split}
\end{equation}
\begin{equation}
\begin{split}
L_D^2 (q_k^1, q_k^2, t) =  \frac{1}{2} \frac{(q_k^2 - q_k^1)^2}{ \nicefrac{h}{2} } + \dots \\ (q_k^2 - q_k^1) \cdot \frac{1}{2} \left ( \vec{a}(q_k^1, \nicefrac{h}{2}) + \vec{a}(q_k^2, h) \right ) - \frac{h}{2}  \varphi(q_k^1, \nicefrac{h}{2})
\end{split}
\end{equation}
\end{subequations}
Because this two-step discrete Lagrangian is the composition of a half step with its adjoint, the resulting Lagrangian is self-adjoint and, hence, $2^{nd}$ order by construction. Taking the DEL equations on these Lagrangians obtains the relations
\begin{equation}
v_k^1 = \frac{q_k^1 - q_k^0}{\nicefrac{h}{2}}
\end{equation}
\begin{equation}
\begin{split}
v_k^1 - v_k^2 + \frac{1}{2}\left ( a(q_k^0, 0) - a(q_k^2, h) \right ) - \dots \\ \frac{1}{2} \nabla_{q_k^1} \left [ (v_k^2 + v_k^1) \cdot  a(q_k^1, \nicefrac{h}{2}) \right ] h - \nabla \varphi (q_k^1, \nicefrac{h}{2}) h = 0
\end{split}
\end{equation}
\begin{equation}
v_k^2 = \frac{q_k^2 - q_k^1}{\nicefrac{h}{2}}
\end{equation}

Here, we note that $q_k^0 = q_k^1 - v_k^1 \nicefrac{h}{2}$ and $q_k^2 = q_k^1 + v_k^2 \nicefrac{h}{2}$, and by Taylor expanding the vector potential, we obtain that
\begin{equation} \label{centralDifference}
\begin{split}
\frac{1}{2} \left ( a(q_k^0, 0) - a(q_k^2, h) \right ) \approx \\ - \frac{1}{2}\left [(v_k^1 + v_k^2) \cdot \nabla \right ] a(q_k^1, \nicefrac{h}{2}) h - \frac{\partial a}{\partial t} (q_k^1, \nicefrac{h}{2}) h + \mathcal{O}(h^3)
\end{split}
\end{equation}
The higher order terms are corrections at the precision of the Lagrangian integral, and therefore does not reduce the order of the integrator scheme. Furthermore, in most electromagnetic particle-in-cell codes, the fields are only second order accurate in the time step anyway. Thus, this approximation does not reduce the precision of the integrator.

By noting that $ \nabla(v \cdot \vec{A}) - (v \cdot \nabla) \vec{A} = v \times \vec{B}$, and defining the normalized magnetic $\vec{b} =  (\nicefrac{e}{m c})\vec{B}$ and electric $\vec{e} = (\nicefrac{e}{m}) \vec{E} $ fields, the update sequence for an explicit magnetic Lagrangian is given by
\begin{equation}
q_k^1 = q_k^0 + v_k^1 (\nicefrac{h}{2})
\end{equation}
\begin{equation}
v_k^2 = v_k^1 + \left ( \frac{1}{2} (v_k^1 + v_k^2 ) \times \vec{b}(q_k^1, t/2) + \vec{e}(q_k^1, t/2) \right ) h
\end{equation}
\begin{equation}
q_k^2 = q_k^1 + v_k^2 (\nicefrac{h}{2})
\end{equation}
This is equivalent to the update sequence proposed by Boris. This proves that the update derived by Boris in 1970 is, in fact, a symplectic integrator.

The reader may note that this Taylor expansion does not necessarily guarantee symplecticity -- could it be possible that we have broken the symplectic two-form by taking an implicit algorithm and making it explicit? This is not the case, as we will now argue. The approximation made earlier may be used as a definition of the electric and magnetic fields, and in fact is invertible. That is to say, we could rewrite the Lagrangian in terms of the electric and magnetic fields to the same order of accuracy as the Lagrangian itself, and re-derive the equations of motion in terms of the fields instead of the potentials. We have included the forward case because it is constructive -- it shows how the fields arise from the potentials in much the same way that they arise from the continuous Lagrangian.

Now we consider the relativistic Lagrangian
\begin{equation}
L(q, \dot{q}, t) = m c^2 \sqrt{ 1 - \nicefrac{\dot{q}^2}{c^2} } + \dot{q} \cdot \frac{e}{c} \vec{A} (q, t) - e \phi(q, t)
\end{equation}
As with the nonrelativistic limit, we consider a split of the form
\begin{subequations}
\begin{equation} \label{relLagrangeA}
\begin{split}
L_D^1(q_k^0, q_k^1, t) = - m c^2 \frac{h}{2} \sqrt{1 -  \left ( \nicefrac{ \frac{q_k^1 - q_k^0}{ \nicefrac{h}{2} }}{c}\right )^2} + \dots \\ (q_k^1 - q_k^0) \cdot \frac{1}{2} \left ( \vec{a}(q_k^0, 0) + \vec{a}(q_k^1, \nicefrac{h}{2}) \right ) - \frac{h}{2} \varphi(q_k^1, \nicefrac{h}{2})
\end{split}
\end{equation}
\begin{equation} \label{relLagrangeB}
\begin{split}
L_D^2 (q_k^1, q_k^2, t) = - m c^2 \frac{h}{2} \sqrt{1 -  \left ( \nicefrac{ \frac{q_k^2 - q_k^1}{ \nicefrac{h}{2} }}{c}\right )^2} + \dots \\ (q_k^2 - q_k^1) \cdot  \frac{1}{2} \left ( \vec{a}(q_k^1, \nicefrac{h}{2}) + \vec{a}(q_k^2, h) \right ) - \frac{h}{2} \varphi(q_k^1, \nicefrac{h}{2})
\end{split}
\end{equation}
\end{subequations}
Employing the same minimization as in the previous section, we obtain the recursion relations
\begin{equation}
q_k^1 = q_k^0 + v_k^1 (\nicefrac{h}{2})
\end{equation}
\begin{equation}\label{relativisticBoris}
\gamma_k^2 v_k^2 = \gamma_k^1 v_k^1 + \left ( \frac{1}{2} (v_k^1 + v_k^2 ) \times \vec{b}(q_k^1, t/2) + \vec{e}(q_k^1, t/2) \right ) h
\end{equation}
\begin{equation}
q_k^2 = q_k^1 + v_k^2 (\nicefrac{h}{2})
\end{equation}

It is necessary to invert equation (\ref{relativisticBoris}) for both $v_k^2$ and $\gamma_k^2$ to obtain the explicit update sequence. This is equivalent to the update method derived by J.-L. Vay \cite{vay:07}, illustrating that the Vay integrator is also symplectic.

\section{Two More Integrators}\label{explicitIntegrator}

An alternative second order Lagrangian is given by
\begin{subequations} \label{shadyGaugeLagrangian}
\begin{equation}
L_D^1 = \frac{1}{2} \frac{(q_k^1 - q_k^0)^2}{\nicefrac{h}{2}} + \left ( q_k^1 - q_k^0 \right ) \cdot a(q_k^1) - \frac{h}{2} \varphi(q_k^1, \nicefrac{h}{2})
\end{equation}
\begin{equation}
L_D^2 = \frac{1}{2} \frac{(q_k^2 - q_k^1)^2}{\nicefrac{h}{2}} + \left ( q_k^2 - q_k^1 \right ) \cdot a(q_k^1) - \frac{h}{2} \varphi(q_k^1, \nicefrac{h}{2})
\end{equation}
\end{subequations}
Again, this is the concatenation of a discrete Lagrangian with its adjoint, and hence is second order. This gives the second order update sequence
\begin{subequations}
\begin{equation}
q_k^1 = q_k^0 + v_k^1 \frac{h}{2}
\end{equation}
\begin{equation}
v_k^2 = v_k^1 + \frac{h}{2} \nabla \left [ \left (v_k^1 + v_k^2 \right ) \cdot a(q_k^1) \right ] - \nabla \varphi(q_k^1, \nicefrac{h}{2}) h
\end{equation}
\begin{equation}
q_k^2 = q_k^1 + v_k^2 \frac{h}{2}
\end{equation}
\end{subequations}
This Lagrangian has the advantage of being second order and directly invertible, but cannot be rewritten in terms of the electric and magnetic fields. Because this may be directly inverted for $v^2_k$ in terms of a matrix inverse, without any sort of root solving, this algorithm can be made explicit through a matrix inversion.

Another Lagrangian that is possible is the implicit midpoint Lagrangian
\begin{equation} \label{midpointLagrangian}
\begin{split}
L_D(q_{k+1}, q_k, h) = \frac{1}{2} \frac{(q_{k+1} - q_k)^2}{h} + \dots \\ (q_{k+1} - q_k) \cdot a\left ( \frac{q_{k+1} + q_k}{2}, \nicefrac{h}{2} \right ) - h \varphi \left ( \frac{q_{k+1} + q_k}{2}, \nicefrac{h}{2} \right )
\end{split}
\end{equation}
The resulting integration scheme, which is second order accurate because it is self-adjoint
\begin{subequations}
\begin{equation} \label{midpoint1}
q_{k} = q_{k-1} + v_{k} h
\end{equation}
\begin{equation} \label{midpoint2}
\begin{split}
v_{k+1} = v_k + a \left ( \frac{q_k + q_{k-1}}{2}, - \nicefrac{h}{2} \right ) - a \left ( \frac{q_{k+1} + q_{k}}{2}, \nicefrac{h}{2} \right ) + \dots \\ h \nabla_{q_k} \biggl \{ v_k \cdot a \left ( \frac{q_k + q_{k-1}}{2}, - \nicefrac{h}{2} \right ) + v_{k+1} \cdot a \left ( \frac{q_{k+1} + q_{k}}{2}, \nicefrac{h}{2} \right ) + \dots \\ - \varphi \left ( \frac{q_{k+1} + q_k}{2}, \nicefrac{h}{2} \right ) - \varphi \left ( \frac{q_k + q_{k-1}}{2}, - \nicefrac{h}{2} \right ) \biggr \}
\end{split}
\end{equation}
\begin{equation} \label{midpoint3}
q_{k+1} = q_{k} + v_{k+1} h
\end{equation}
\end{subequations}
where the implicit dependence on the coordinates can be replaced with velocity by substituting eqn. (\ref{midpoint3}) and eqn. (\ref{midpoint1}) into eqn. (\ref{midpoint2}). We will consider this update sequence further in the next section.

\begin{figure} \label{orderOfIntegrator}
\includegraphics[scale=0.75]{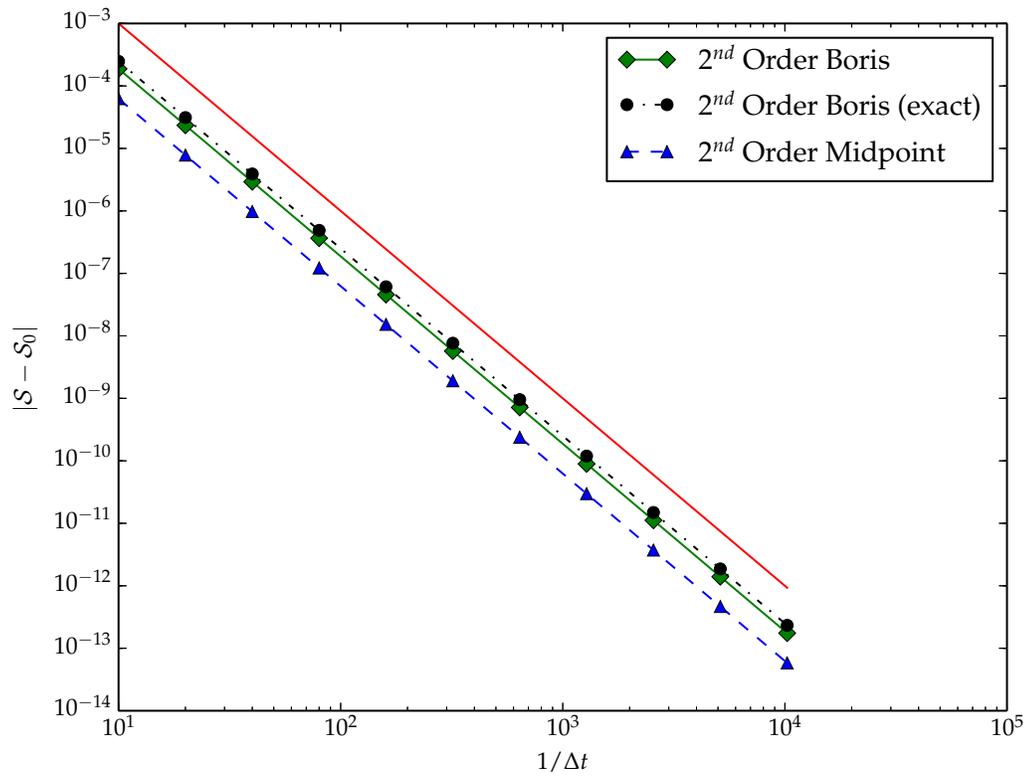}
\caption{A comparison of the order of convergence of the action integral for this system. The red line is $\Delta t^{-3}$.}
\end{figure}

In figure (\ref{orderOfIntegrator}) we consider the order of convergence of the action integral for a system with a constant magnetic field. As per Marsden and West~\cite{marsden_west:01}, the order of an integrator is given by the order of the correction to the exact integral
\begin{equation}
\int_0^h L(\dot{q}, q, t) dt = \sum_i L^i_D(q_k^i, q_k^{i+1}, \gamma_i h) + \mathcal{O}(h^{n+1})
\end{equation}
for an order-$n$ symplectic integrator. 

In figure \ref{orderOfIntegrator}, we consider the order of three of the integration schemes: the Boris integrator, the exact Boris Lagrangian (action computed using the same Lagrangian), and the implicit midpoint. As is clear, all of the integrators are second order symplectic integrators which give comparable approximations to the action integral.  The choice of integrator will depend upon the specific application.

In the next section, we will see an issue with higher order integrators.

\section{Higher Order Integration Schemes}

The discrete Lagrangians with the vector potential are self-adjoint (see \S 2.4 of \cite{marsden_west:01}), and therefore one might assume the composition map that follows the pattern due to Yoshida~\cite{yoshida:90} will generate higher order integrators. This is not the case, however.

There is a subtlety to the Yoshida trick. At one point during the derivation, it is assumed that the inverse of the $2n^{th}$ order map $\mathcal{M}^{2n}(t)^{-1} = \mathcal{M}^{2n}(-t)$. That is, the dynamics can be reversed simply by running backwards in time. Each one of these integrators has that property, and thus the Lagrangian taken by summing the individual Lagrangians that generate the second order maps, with the proper time steps, will yield fourth order integrators, viz.
\begin{equation}
L_D^{(4)} = L_D^{(2)}(\gamma h) + L_D^{(2)}((1 - 2 \gamma)h) + L_D^{(2)}(\gamma h)
\end{equation}

However, we must be careful in the application of these methods. For one, the central differencing using to obtain the electric and magnetic fields from the potential fields in the derivation of the Boris and Vay integrators are only second order accurate in time. It is not clear that this happy circumstance would occur again for fourth order, and hence these integrators do not immediately generalize to fourth order. The exact update, however, is fourth order.

This has further implications for self-consistent electromagnetic algorithms. The canonical approach of mixing the Boris/Vay integrators with a staggered Yee mesh~\cite{yee:66} computation of the fields is only second order accurate in the time step. A fourth order integrator would require a fourth order central differencing plus a fourth order staggered mesh Such a simulation would be extremely memory intensive for the number of past time steps required for a fourth order central differencing.

Finally, such an integrator would have to have the correct time centering on the fields, which may prove difficult to do in conjunction with the other requirements, such as charge conservation. There is, at present, no algorithm for a fourth order charge conservation scheme analogous to the Esirkepov scheme~\cite{esirkepov:2001} for second order algorithms It is therefore unlikely that a fourth order self-consistent electromagnetic particle-in-cell algorithm is practical.

\section{Conclusion}

We have presented four integration schemes for magnetic systems and shown that the Boris and Vay integrators in the literature are derivable from a discretized Lagrangian, and hence symplectic. Furthermore, each integrator that uses the potentials directly, without the Taylor expansion required to obtain the Boris and Vay integrators, are inverted by a simple time reversal, and hence are subject to the Yoshida method for generating higher order integrators.

\section{Acknowledgements}

The author would like to thank Dan T. Abell, Kevin Paul, Peter Stoltz, and John Cary (Tech-X) for helpful discussions. This work has been paid for by Tech-X Corporation, and by the Department of Energy, Office of Science, under grant no. DE-SC0004432.

\bibliographystyle{elsarticle-num}
\bibliography{symplecticMagnetic.bib}






% \end{thebibliography}

\end{document}